\documentstyle[12pt]{article}

\begin{document}

\begin{flushright}
IMSc/2014/12/14
\end{flushright} 

\vspace{2mm}

\vspace{2ex}

\begin{center}

{\large \bf Singularity Resolution + Unitary } \\

\vspace{2ex}

{\large \bf Evolution + Horizon = Firewall ?} \\

\vspace{6ex}

{\large  S. Kalyana Rama}

\vspace{3ex}

Institute of Mathematical Sciences, C. I. T. Campus, 

Tharamani, CHENNAI 600 113, India. 

\vspace{1ex}

email: krama@imsc.res.in \\ 

\end{center}

\vspace{6ex}

\centerline{ABSTRACT}
\begin{quote} 

We assume that a quantum gravity theory exists where evolutions
are unitary, no information is lost, singularities are resolved,
and horizons form. Thus a massive star will collapse to a black
hole having a horizon and an interior singularity--resolved
region. Based on unitarity and on a postulated relation, we
obtain an evolution equation for the size of this region. As the
black hole evolves by evaporation and accretions, this region
grows, meets the horizon, and becomes accessible to an outside
observer. The required time is typically of the order of black
hole evaporation time. If the boundary of the
singularity--resolved region can be considered as the firewall
of Almheiri et al then this marks the appearance of a firewall.

\end{quote}

\vspace{2ex}









\newpage


\vspace{4ex}

{\bf (1)}
Consider the gravitational collapse of a very massive star. Such
a collapse is believed to result in the formation of a black
hole, containing a singularity at the center and a horizon
surrounding it, with the constituents of the star taken to have
disappeared down the singularity, and with the spacetime outside
the horizon being given, for example, by the Schwarzschild
solution in the static spherically symmetric case. The black
hole horizon has Bekenstein--Hawking entropy and emits Hawking
radiation until the horizon reaches Planckian size. As Hawking
has argued, such an evolution is non unitary and leads to
information loss.

In any quantum gravity theory, the evolutions are expected to be
unitary, no information is expected to be lost, and the
singularities are expected to be resolved. However, whether
horizons are present or absent is still debatable. For a sample
of works containing similar ideas regarding singularity and/or
unitarity and/or horizon, see \cite{morgan} -- \cite{gp} where
horizons are assumed to be present, \cite{fuzz} for Mathur's
fuzzball proposal which postulates that horizons are absent,
\cite{page} where high entropic objects are considered, and
\cite{k12} where it is argued that horizons are absent.

In this paper, we assume that a quantum gravity theory exists
where evolutions are unitary, no information is lost, and
singularities are resolved. We further assume that horizons are
present, atleast initially, having the same properties as found
by standard semi classical analysis. Thus, in such a theory, the
gravitational collapse of a very massive star will result in the
formation of a black hole which now contains a
singularity--resolved region at the center, and a horizon
surrounding it which has an entropy given by Bekenstein--Hawking
formula and emits Hawking radiation. The spacetime outside the
horizon will be given by standard solutions in general
relativity. In the following, for the sake of brevity, we refer
to the singularity--resolved region at the center as singularity
cloud.

We consider a spherically symmetric black hole, and analyse the
consequences of the singularity resolutions and unitary
evolutions on the singularity cloud and on the horizon as the
black hole evolves further by evaporation and any possible
accretions. Postulating a simple relation between the size of
the singularity cloud and its contents, and based on the
unitarity of the evolution, we obtain an equation for the size
and analyse its consequences. 

We find that, much before its size becomes Planckian, horizon
will meet the singularity cloud. The singularity cloud then
becomes accessible to an outside observer, thus the horizon
ceases to exist. The boundary of this singularity cloud can
perhaps be considered as the firewall of Almheiri et al
\cite{amps}. If so then access to the singularity cloud marks
the appearance of firewall. The time of this appearance depends
on the assumptions made about the size of the singularity cloud,
and also on any accretions that may take place in the
intervening period, but it is typically of the order of black
hole evaporation time. 

We conclude with a summary and a few questions for further
studies.


\vspace{4ex}

{\bf (2)}
Consider a spherically symmetric black hole formed in a stellar
collapse in $d + 1$ dimensional spacetime. Let $R_h, S, M$, and
$T$ denote its horizon radius, entropy, mass, and Hawking
temperature respectively.  In Planckian units, $S, M$, and $T$
may be written in terms of $R_h$ as
\begin{equation}\label{smt}
S = C_S \; R_h^{d - 1} \; \; \; , \; \; \; \; 
M = C_M \; R_h^{d - 2} \; \; \; , \; \; \; \; 
T = C_T \; R_h^{- 1} 
\end{equation}
where $C_S$ and $C_M$ are positive numerical constants and $C_T
= \frac {d - 2} {d - 1} \; \frac{C_M} {C_S} \;$. Let the initial
values of $(R_h, S, M)$ be given by
\begin{equation}\label{ic0}
(R_h, S, M) \vert_{t = 0} \; = \; (R_{h 0}, S_0, M_0) \; \gg 1
\end{equation}
where $(R_{h 0}, S_0, M_0)$ obey the relations given in equation
(\ref{smt}) and $t = 0$ is taken as the black hole formation
time, namely as the time when the transients of the collapse
have died down and the black hole has begun to emit Hawking
radiation.

The black hole evaporates by emitting Hawking radiation,
therefore its mass will decrease. Also, the black hole may
accrete more matter in which case its mass will increase. Thus,
$(R_h, S, M, T)$ are all functions of time. The rate of mass
loss due to evaporation $\propto \; (Area) \; (temperature)^{d +
1}$ and, hence, one obtains $\dot{M}_{evap} = - \; \frac
{C_{evap}} {R_h^2} \;$ where $C_{evap}$ is a positive numerical
constant and an overdot denotes time derivative.  Similarly, the
rate of mass gain due to accretion may be written as
$\dot{M}_{accr} = g(t) \; \ge 0 \;$. The total rate of mass
change is, therefore, given by 
\begin{equation}\label{mdot}
\dot{M} \; = \; \dot{M}_{evap} + \dot{M}_{accr} \; = \; 
- \; \frac{C_{evap}}{R_h^2} \; + \; g(t) \; \; . 
\end{equation}
The corresponding changes in the entropy $S$ are obtained by
$\delta S = \frac {\delta M} {T} \;$. The total rate of entropy
change is, therefore, given by
\begin{equation}\label{sdot}
\dot{S} \; = \; \dot{S}_{evap} + \dot{S}_{accr} \; = \;
\frac{R_h}{C_T} \; \left( - \; \frac{C_{evap}}{R_h^2} \;
+ \; g(t) \right) \; \; . 
\end{equation}
Equations (\ref{smt}) and (\ref{mdot}), or (\ref{sdot}), lead to
an equation for $R_h(t)$ given by
\begin{equation}\label{rhdot}
\dot{R}_h \; = \; \frac{R_h^{3 - d}}{(d - 2) \; C_M} \; 
\left( - \; \frac{C_{evap}}{R_h^2} \; + \; g(t) \right) \; \; ,
\end{equation}
which is to be solved with an initial condition $R_h(0) = R_{h
0} \;$. Equations (\ref{smt}) then give the entropy $S(t)$ and
the mass $M(t) \;$.


\vspace{4ex}

{\bf (3)}
Consider the interior of the black hole. In the standard
semiclassical analysis, there is a singularity of Planckian size
at the center and the constituents of the collapsing star are
taken to have disappeared down the singularity. Now, we are
assuming that a quantum gravity theory exists where the
singularities are resolved, evolutions are unitary, and no
information is lost. Hence, in the interior of the black hole
now, there should be no singularities. Furthermore, whatever is
in the interior must be capable of encoding the information
about the black hole formation and its further evolution. 

Let the singularity be resolved, and let the resulting
singularity--resolved region be of size $L$ and contain $N$
number of `quantum gravity units' which can encode information
of order one bit each. For the sake of brevity, we refer to the
singularity--resolved region as singularity cloud, and `quantum
gravity units' as simply units. The size $L$ and the number of
units $N$ will depend on time $t$ as the black hole evolves
further by evaporation and accretions.

It is physically reasonable to expect, and hence we assume, that
the initial number of units $N_0$ equals the initial entropy
$S_0$ of the black hole formed, so that information about the
collapse of the star and the formation of black hole can all be
encoded in the singularity cloud. Also, in a theory where the
singularities are resolved, the size $L$ should depend on $N \;$
and be parametrically much larger than Planck length. Otherwise
there will be `remnant problems'. Thus we write $L \propto
N^\sigma$ and, most conservatively, require only that $\sigma >
0 \;$. Note that: {\bf (i)} Semiclassically $L = {\cal O} (1)$,
equivalently $\sigma = 0 \;$. {\bf (ii)} The value $\sigma =
\frac{1}{d}$ is natural since it implies a density of order one
unit per unit Planck volume. {\bf (iii)} $\sigma > \frac{1}{d -
1}$ implies that the initial size of the singularity cloud $L_0$
is larger than the initial horizon size $R_{h 0}$ which means
that there is no horizon. Hence $\sigma \le \frac{1}{d - 1} \;$
since horizon is assumed to be present atleast initially. Thus,
similarly as in equation (\ref{smt}), we write the relation
between the size $L$ of the singularity cloud and the number of
units $N$ contained in it as

\begin{equation}\label{l}
N = C_N \; L^\alpha \; \; \; , \; \; \; \;
d - 1 \le \alpha < \infty 
\end{equation}
where $C_N$ is a positive numerical constant and $\alpha = \frac
{1} {\sigma} \;$. The ratio $\frac{L}{R_h}$ is given by 
\begin{equation}\label{lr}
\left( \frac{L}{R_h} \right)^\alpha \; = \; 
\frac {C_s^{\frac{\alpha}{d - 1}}} {C_N} \; \left(
\frac {N} {S^{\frac{\alpha}{d - 1}}}  \; \right) \; \; . 
\end{equation}
The assumption that horizon is present initially implies that
$L_0 < R_{h 0} \;$. Since $N_0 = S_0 \gg 1 \;$, it then follows
that $\alpha \ge d - 1$ and $C_N > C_S$ if $\alpha = d - 1 \;$.

Now consider the evolution of the number $N$ as the black hole
evaporates and accretes. The number $N$ encodes the information
about evolution. Hence, the rate of change of $N$ due to
accretion should be given by
\[
\dot{N}_{accr} =  \dot{S}_{accr} =
\frac {\dot{M}_{accr}} {T} \; \ge 0 
\]
so that accretion information can be encoded in the singularity
cloud.

Consider evaporation, due to which the entropy and mass of the
black hole decrease. If $\dot{N}_{evap} = \dot{S}_{evap}$ then
$\dot{N}_{evap} < 0$ and, eventually, these decreasing number of
units will not be able to encode the information about the
formation and the evolution of the black hole. However, if
$\delta N = \vert \delta S \vert$ in any process where the
entropy $S$ of the black hole is changed then $\delta N > 0$ and
the information about such a change can continue to be encoded
in the singularity cloud by these increasing number of units. We
assume that this is the case. Then the rate of change of $N$ due
to evaporation should be given by

\[
\dot{N}_{evap} =  \vert \dot{S}_{evap} \vert \; > 0 \; \; , 
\]
which may also be thought of as the ingoing Hawking photons
giving rise to new quantum gravity degrees of freedom upon
falling into the singularity cloud. Thus, the total rate of
change of the number of units $N$ in the singularity cloud is
given by

\begin{equation}\label{ndot}
\dot{N} \; = \; \dot{N}_{evap} + \dot{N}_{accr} \; = \;
\frac{R_h}{C_T} \; \left( \frac{C_{evap}}{R_h^2} \;
+ \; g(t) \right) \; > 0 \; \; . 
\end{equation}
The corresponding change in the size $L$ then follows from
equation (\ref{l}).

Thus, for given initial values and a given accretion rate, one
solves equation (\ref{rhdot}) for $R_h(t) \;$. One then solves
equation (\ref{ndot}) for $N(t) \;$. Equations (\ref{smt}) and
(\ref{l}) then give $S$, $\; M$, and $L$ as functions of $t
\;$. Note that $N$ and $L$ increase with respect to $t \;$, and
that $R_h$ decreases if the accretion rate is sufficiently
small. Thus, once the accretion rate has become sufficiently
small or vanished for good, there will eventually be a time $t =
t_{fw}$ when $L = R_h \;$. This is when the singularity cloud
meets horizon and becomes accessible to an outside observer. If
the boundary of the singularity cloud can be taken as the
firewall of Almheiri et al then $t_{fw}$ is the time when the
firewall appears.


\vspace{4ex}

{\bf (4)}
Consider equation (\ref{rhdot}) for $R_h(t) \;$. It is a non
autonomous, first order, ordinary differential equation. For a
given arbitrary function $g(t)$, it is reasonably easy to
understand the behaviour of the solution but, in general, it is
not possible to obtain a closed form expression for $R_h(t)
\;$. Hence, we first present a few illustrative examples. In the
following, let $t_{fw}$ be the time when the singularity cloud
meets the horizon. Then
\begin{equation}\label{lrsfw}
L_{fw} = R_{h \; fw} = \left( \frac {S_{fw}} {C_s}
\right)^{\frac{1}{d - 1}} 
\end{equation}
where the subscripts $fw$ on the variables denote their values
at $t = t_{fw} \;$. 

\vspace{4ex}

{\bf Example (a) :} 
Let there be no accretion, {\em i.e.} $g(t) = 0 \;$. The
solutions are then given by $N = 2 S_0 - S$ and
\begin{equation}\label{aexample}
R_h = R_{h 0} \; \left( 1 - \frac{t}{t_e} \right)^{\frac{1}{d}}
\; \; , \; \; \; \;
L = L_0 \; \left( 2 - \left( 1 - \frac{t}{t_e}
\right)^{\frac{d - 1}{d}} \right)^{\frac{1}{\alpha}}
\end{equation}
where the evaporation time $t_e$ is given by 
\begin{equation}\label{tevap}
t_e = \frac {d - 2} {d} \left( \frac{C_M} {C_{evap}} \right)
R_{h 0}^d 
\end{equation} 
and $S$ and $M$ can be obtained from equations (\ref{smt}).
Clearly $R_h$ decreases and $L$ increases, becoming equal at $t
= t_{fw} \;$. Then, using $N = 2 S_0 - S$, it can be seen from
equation (\ref{lr}) that if $\alpha = d - 1$ then
\begin{equation}\label{astfw}
S_{fw} = \frac {2 \; C_S} {C_S + C_N} \; S_0 
\; \; , \; \; \;  \; \; \;  
t_{fw} = \left(1 - \frac {2 \; C_S} {C_S + C_N}
\right)^{\frac {d} {d - 1}} \; t_e \; \; 
\end{equation}
whereas if $\alpha > d - 1$ then $S_{fw} \propto S_0^{\frac {d -
1} {\alpha}} \ll S_0$ and $t_{fw} \sim t_e \;$. Also, 
\[
L_{fw} = R_{h \; fw} = \left( \frac {S_{fw}} {C_s}
\right)^{\frac{1}{d - 1}} \; \sim \; 
R_{h 0}^{\frac {d - 1} {\alpha}} \; \gg 1 \; \; .  
\]

\vspace{4ex}

{\bf Example (b) :} 
Let the accretion equal the initial evaporation, {\em i.e.}
$g(t) = \frac {C_{evap}} {R_{h 0}^2} \;$. The solutions are then
given by $(R_h, S, M) = (R_{h 0}, S_0, M_0)$ and
\begin{equation}\label{bexample}
N \; = \; S_0 \; \left( 1 + \frac{2 (d - 1)}{d} \frac{t}{t_e}
\right)
\; \; , \; \; \; \; 
L \; = \; L_0 \; \left( 1 + \frac{2 (d - 1)}{d} \frac{t}{t_e}
\right)^{\frac{1}{\alpha}}
\end{equation}
where $t_e$ is defined in equation (\ref{tevap}). Clearly $L$
increases and becomes equal to $R_h$ at $t = t_{fw} \;$. Then
$L_{fw} = R_{h 0}$ and $t_{fw}$ is given by 
\begin{equation}
t_{fw} \; = \; \frac {d} {2 (d - 1)} \; \left( \frac {C_N} {C_S}
\; R_{h 0}^{\alpha - (d - 1)} - 1 \right) \; {t_e} 
\; \; \sim R_{h 0}^{\alpha - (d - 1)} \; t_e \; \; . 
\end{equation}
Thus, $t_{fw} \sim t_e$ if $\alpha = d - 1$ and $t_{fw}$ is
parametrically much larger than $t_e$ if $\alpha > d - 1 \;$.

\vspace{4ex}

{\bf Example (c) :} 
Let there be no evaporation, {\em i.e.} $C_{evap} = 0 \;$. The
solutions are then given by $N = S = C_S \; R_h^{d - 1}$ and 
\begin{equation}\label{cexample}
R_h = \left( \frac {M} {C_M} \right)^{\frac{1}{d - 2}}
\; \; , \; \; \; \; 
M = M_0 + \int_0^t d t \; g(t) 
\end{equation}
where $g(t) \ge 0 \;$. Hence, $R_h$, $\; S$, and $M$ are non
decreasing. Since $N = S$, we also have
\[
\left( \frac{L}{R_h} \right)^\alpha \; = \; 
\left( \frac{L_0}{R_{h 0}} \right)^\alpha \; 
\left( \frac{R_h} {R_{h 0}} \right)^{d - 1 - \alpha} \; \; . 
\]
Since $L_0 < R_{h 0}$ by assumption, $R_h \ge R_{h 0}$, and $d -
1 \le \alpha \;$, it follows that $L$ remains less than $R_h \;$
and, hence, the singularity cloud does not meet the horizon. 


\vspace{4ex}

{\bf General Case :}
Consider the general case where $g(t)$ is a non negative
arbitrary function. It can be seen from equation (\ref{rhdot})
that if $g(t)$ is less than the evaporation term $\frac
{C_{evap}} {R_h^2}$ then $\dot{R}_h$ is negative, $R_h$
decreases, $\frac{C_{evap}} {R_h^2}$ increases, and $g(t)$
becomes even less important. The solution in such a case is of
the type given in example {\bf (a)}: $R_h$ decreases, $L$
increases, and the singularity cloud and the horizon meet at a
time $t_{fw}$ which is of the order of evaporation time given in
equation (\ref{tevap}). Similarly, if $g(t)$ is greater than the
evaporation term $\frac{C_{evap}} {R_h^2}$ then $\dot{R}_h$ is
positive, $R_h$ increases, $\frac{C_{evap}} {R_h^2}$ decreases,
and $g(t)$ becomes even more important. The solution in such a
case is of the type given in example {\bf (c)}: Both $R_h$ and
$L$ increase, but the singularity cloud and the horizon move
apart from each other and do not meet.

Note also that for a given value of $R_h$, the values of $S$ and
$M$ are uniquely given by equation (\ref{smt}). However, unless
the accretion is always absent as in example {\bf (a)}, the
values of $L$ and $N$ are not unique. This non uniqueness of $L$
and $N$ for a given value $R_h$ can be seen in example {\bf
(b)}. In general, $L$ and $N$ depend on the details of the black
hole evolution from the time of its formation.

Putting together these properties, one can see the following
features of the solutions of equations (\ref{rhdot}), (\ref{l}),
and (\ref{ndot}). They can also be proved more rigorously.

\begin{itemize}

\item 

The singularity cloud and the horizon do not meet during the 
accretion phase.

\item

They can meet only during the evaporation phase.

\item

The time required for this meeting is {\em generically} of the
order of evaporation time given in equation (\ref{tevap}), but
now with $R_{h 0}^d$ replaced by $R_{h \; max}^d$ where $R_{h \;
max}$ is the maximum horizon size reached due to various
accretions. However, as can be seen in example {\bf (b)}, if the
model parameter $\alpha > d - 1$ then it is possible to increase
this time arbitrarily by fine tuning the accretion rate.

\item

The black hole quantities, such as its entropy $S$ and mass $M$,
are uniquely fixed by its horizon radius $R_h \;$. However, the
singularity cloud quantities, such as its size $L$ and the
number of units $N$, are not uniquely fixed by $R_h \;$ but 
depend on the details of the black hole evolution.

In particular, knowing $R_h$ at a given time $t$ alone is not
sufficient to predict when in future the singularity cloud will
meet the horizon and become accessible to an outside observer.


\end{itemize}

The unpredictability of $L$ and $N$, and hence of the time
$t_{fw}$ when the singularity cloud meets the horizon, is due to
the presence of the horizon and the accretions onto the black
hole. This has been referred to as `loss of predictability' in
our earlier works \cite{k12} and has been used to argue that
horizons should be absent.


\vspace{4ex}

{\bf (5)} 
In summary: We have assumed that a quantum gravity theory exists
where evolutions are unitary, no information is lost, and
singularities are resolved. In a gravitational collapse of a
very massive star in such a theory, we further assumed that
horizons are present, atleast initially, having the same semi
classical properties. A black hole in such a theory has a
singularity--resolved region at the center, referred to as
singularity cloud, which is of size $L$ and which contains $N$
number of `quantum gravity units' so that information about the
formation and further evolution of the black hole can be
encoded.

We considered a spherically symmetric black hole. Its horizon
size $R_h$ decreases due to Hawking radiation and increases due
to accretions. An equation for $R_h(t)$ is easy to write. For
the singularity cloud, its size $L$ should be parametrically
large and should depend on the number of units $N \;$.
Postulating a simple relation between $L$ and $N$, and based on
the unitarity of the evolution, we then obtained an equation for
$N(t)$ and thus for $L(t)$ also.

Analysing the resulting equations for $R_h$ and $N$, we found
that, generically as the black hole evolves, the singularity
cloud increases in size, meets the horizon at time $t_{fw}$, and
becomes accessible to an outside observer. The horizon thus
ceases to exist. If the boundary of the singularity cloud can be
considered as the firewall of Almheiri et al then the time
$t_{fw} \;$ marks the appearance of the firewall. The time
$t_{fw} \;$ depends on the details of the evolution but is
typically of the order of black hole evaporation time.

Also, knowing $R_h$ at a given time $t$ is sufficient to predict
the black hole quantities such as its entropy and mass. However,
this alone is not sufficient to predict the singularity cloud
quantities, such as its size $L$ or the time of access to an
outsider $t_{fw}$, since they depend also on the details of the
black hole evolution.

\vspace{1ex}

We now conclude by posing a few questions for further studies.

\vspace{1ex}

Can the boundary of the singularity cloud be taken as the
firewall of Almheiri et al? Does its appearance to an outsider
at time $t_{fw}$ resolve the issues raised in their paper
\cite{amps}?

\vspace{1ex}

When there are accretions, knowing $R_h$ alone at a given time
is not sufficient to predict $t_{fw} \;$. Since, by assumption,
spacetime outside the horizon is described as in standard
semiclassical analysis, there seems to be no way at all to
predict $t_{fw}$ except by knowing all the past evaporation and
accretions onto the black hole since the time of its formation.
Is such an unpredictability desireable, or does it point to some
inconsistencies? Our own view, see \cite{k12}, is that this
unpredictability is undesireable, and implies that horizons do
not form.

\vspace{1ex}

Once the singularity cloud becomes accessible to an outside
observer, namely for $t > t_{fw}$, its further evolution is
likely to be that of a standard quantum system with large number
of interacting degrees of freedom. What is the nature of these
degrees of freedom and the interactions between them? A related
question is: how does the singularity cloud form in the
gravitational collapse of a very massive star?

\vspace{4ex}


{\bf Note added :} J. Pullin informed us that reference
\cite{gp} contains a concrete implementation of the ideas
presented here. We thank him for the information.


\end{document}